\documentclass[useAMS, usenatbib]{mn2e} 
\usepackage{graphicx}
\usepackage{aas_macros}
\title[Dynamical Opacity-Sampling Models of Mira Variables]
{Dynamical Opacity-Sampling Models of Mira Variables. I:
  Modelling Description and Analysis of Approximations.}  

\author[M.J. Ireland et al.]{M.J. Ireland$^{1,2}$, M. Scholz$^{1,3}$, P.R. Wood$^4$ \\
$^1$School of Physics, University of Sydney NSW 2006, Australia\\
$^2$Division of Geological and Planetary Sciences, MC 150-21, California Institute of
  Technology, 1200 E. California Blvd., Pasadena, CA 91125, USA\\
$^3$Zentrum f\"ur Astronomie der Universit\"at Heidelberg (ZAH),
  Institut f\"ur Theoretische Astrophysik, Albert-Ueberle-Str.2, 69120 Heidelberg, Germany\\
$^4$Research School for Astronomy and Astrophysics, Australian National 
University, Canberra ACT 2600, Australia}

\begin{document}

\pagerange{\pageref{firstpage}--\pageref{lastpage}} \pubyear{2008}

\maketitle

\label{firstpage}

\begin{abstract}
 We describe the Cool Opacity-sampling Dynamic EXtended ({\tt CODEX})
 atmosphere models of Mira
 variable stars, and examine in detail the physical and numerical
 approximations that go in to the model creation. 
 The CODEX atmospheric models are obtained by computing the 
 temperature and the chemical and radiative states of the atmospheric layers, 
 assuming gas pressure and velocity profiles from Mira pulsation models,
 which extend from near the H-burning shell to the outer layers of the atmosphere.  
 Although the code uses the
 approximation of Local Thermodynamic Equilibrium (LTE)
 and a grey approximation in the dynamical
 atmosphere code, many key observable quantities, such as infrared
 diameters and low-resolution spectra, are predicted robustly in spite of these
 approximations.  We show that in visible light,
 radiation from Mira variables is dominated by fluorescence scattering
 processes, and that the LTE approximation likely under-predicts
 visible-band fluxes by a factor of two.
\end{abstract}

\section{Introduction} 

Mira-type variable stars represent the last easily observable stage
in the evolution of solar-mass stars. Due to the interactions between
pulsation, shocks, complex chemistry and radiation pressure, the environment
between the continuum-forming photosphere and the dusty wind is complex
and difficult to model. For this reason, it has not been possible to
extract fundamental parameters of Mira variables (e.g. mass,
metallicity and mass-loss rate) from their spectra and light curves alone.
Similarly, there are many easily observable properties of Mira
variables (such as their visible light curves) that 
have yet to be put in a firm theoretical context. This is in contrast
to Cepheid and RR Lyrae variables, where non-linear effects observable
in their light curves have been used to derive accurate (but
model-dependent) masses and distances
\citep{Keller06,Marconi05}. 

Furthermore, there has been little detailed theoretical work on the
regions of M-type Mira atmospheres 
between the continuum forming photosphere and the radii approximately
10 times more distant where standard \citep{Draine84} dust types are
stable. These layers are particularly crucial for understanding the
interaction between pulsation and mass loss for Mira variables
\citep{Wood79,Bowen88,Hofner98}. 

There is a small set of models from other groups that include 
the effects of interior pulsation by introducing an artificial piston
at the base of the atmosphere, although luminosity variations at this
position are ignored. These calculations have produced physically
consistent dynamical models of the photosphere combined with
self-consistent chemical equilibrium calculations and
radiative transfer. For C-type Mira variables, this kind of
literature is most extensive, with the paper series beginning with
\citet{Hofner98} now including a full treatment of homogeneous
nucleation and non-grey radiative transfer as part of the 1-D
dynamical atmosphere code. 
 
For M-type Mira variables, models are faced with a more complex
chemistry and dust formation, and at least for Miras with periods less
than about 500 days, chaotic motions of the atmosphere play a
larger role in dynamics than the winds resulting from radiative
acceleration of dust grains. Several 
approaches have been attempted. The models of \citet{Hofner03} use a
non-grey radiative transfer in their dynamical models with mean opacities 
in 51 frequency meshes, do not consider the formation of dust grains
in detail and have not yet been compared with observations. The models
of \citet{Jeong03} use a grey approximation for molecular opacities and a
composition-independent dust opacity, but have a sophisticated
treatment of the dust nucleation processes. This approach is more
applicable to the longer period very dusty Mira variable they modelled
than to the more common Miras with periods less than 500 days. 

There are no 3-D models of Mira variables at this
time, but 2-D models of C-rich Mira variables by \citet{Woitke06}
show the key properties of the shock-dominated dynamics: chaos over
large spatial scales similar to the cycle-to-cycle variations seen in
1-D models, and fine structure in the shock fronts caused by the
Rayleigh-Taylor instability.

The models in this paper are an improvement on the \citet{Hofmann98} modelling
scheme. Those models were one-dimensional and were based on a three-step
modelling process to optimize computational efficiency. First, a grey
dynamical model was constructed in which pulsation was self-excited: 
i.e. there was no `piston' artificially causing the pulsation,
and luminosity variations arising in the interior were accounted for.
Second, a non-grey radiative transfer scheme was used to
calculated the detailed temperature profile of the
atmosphere. Finally, more detailed integration through the final
atmospheric profile enabled spectra and center-to-limb profiles to be
calculated. 

Here we describe our Cool Opacity-sampling Dynamic EXtended ({\tt
  CODEX}) atmospheric model series for M-type
(oxygen-rich) Mira variables. The models include
self-excited pulsation with new approximations for convective energy
transport \citep{Keller06}, and an opacity-sampling method for radiative
transfer in LTE. These models are described in detail in
Section~\ref{sectMethods}. They are tested first in
Section~\ref{sectQuality} by further 
calculations that examine the errors in the approximations caused by
our three-step modelling process, and some exploratory calculations
into the effects of the LTE
approximation and the effect of the dynamical atmosphere
  (i.e. velocity stratification) on 
the radiative transfer. The models are further tested in
Section~\ref{sectObs} by comparison with
spectroscopy and optical interferometry, with the aim of determining
which model predictions are expected to be most reliable.

\section{Modelling Assumptions and Methods}
\label{sectMethods}

\subsection{The pulsation models}
\label{sectPulsation}

The aim of the pulsation models was to produce a Mira model with a
period of close to 330 days, which 
matches the period of the local Mira variable $o$~Ceti.  The
near-infrared JHK photometry of \citet{Whitelock00}, combined with the
distance estimate of 107 pc to $o$~Ceti from the Hipparcos parallax
\citep{Knapp03} or the LMC ($M_{\rm bol}$, $\log P$) relation for a 330 day
Mira variable \citep{Hughes90,Feast89}, yields a mean
luminosity of close to 5400 L$_{\odot}$.  This mean luminosity was adopted for
the model: giving it the name {\em o54}.  Given the luminosity, the
luminosity-core mass relation 
\citep[e.g][]{WZ81} defines the mass of the core (0.568 M$_{\odot}$)
below the inner boundary of the pulsating envelope, which is set at about
10$^{7}$ K and a radius of 0.3 R$_{\odot}$.  A mass of 1.1 M$_{\odot}$ was
adopted, to match the mass estimated from Galactic kinematics for a Mira of
period $\sim$330 days \citep{Jura92}.  A metal abundance Z = 0.02 was
adopted, along with a helium abundance Y=0.3, which is close to the envelope
helium abundance of a 1.1 M$_{\odot}$ star on the AGB after it has undergone
first and second dredge-up \citep{Bressan93}.

The self-excited pulsation models were made with the pulsation code described
in \citet{Keller06}.  Given the physical input parameters above, and
adopted values of the model parameters $\alpha_m$ (mixing-length in units of pressure
scale heights) and $\alpha_{\nu}$ (the turbulent viscosity parameter)
the model {\em naturally pulsates}, with the
amplitude limited by the non-linear loss processes of shocks in
the outer atmosphere and turbulent viscosity.  After beginning with
a static model, the nonlinear pulsation model was run to limiting
amplitude.  The period of this
model was then compared to the desired period of 330 days, and the amplitude
was compared to the observed pulsation amplitude (from the photometry of
Whitelock et al . 2000).  The value of $\alpha_m$ was then adjusted until the
correct period was obtained, and $\alpha_{\nu}$ was adjusted to give the
correct pulsation amplitude.  The final values adopted were $\alpha_m$ = 3.5 and
$\alpha_{\nu}$ = 0.25.  The static model of the so-called parent star
with 5400 L$_{\odot}$ has a photospheric radius $R_p$ = 215
R$_{\odot}$ and an effective temperature $T_{\rm eff}$ = 3380 K. The
pulsation period of this model in the linear approximation was 330 days.

\begin{figure}
\hspace{-5mm}
\includegraphics[width={1.05\columnwidth}]{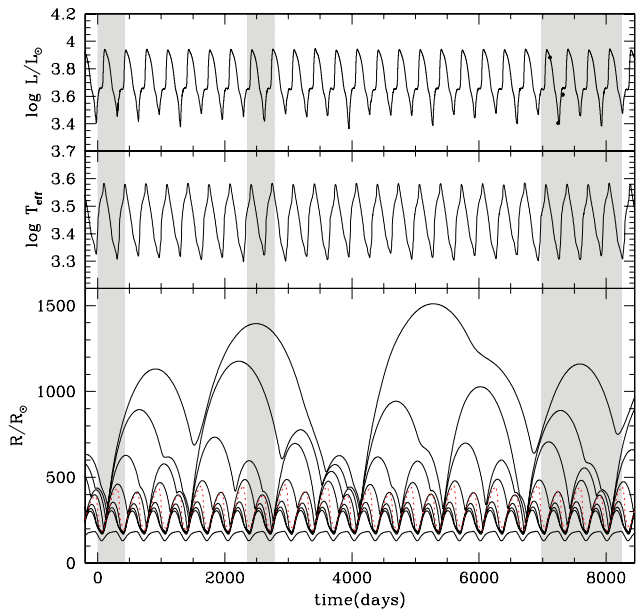}
 \caption{The luminosity (top panel), effective temperature (center
   panel) and radius of selected mass zones (lower panel) as a
   function of time for the {\em o54} pulsation model. The red dashed
   line in the bottom panel corresponds to the radius at Rosseland
   mean optical depth 2/3, and the effective temperature ($\propto
   (L/R^2)^{0.25}$) refers to this radius.}
 \label{figo54Wood}
\end{figure}

The long-term behavior of these models is illustrated in
Figure~\ref{figo54Wood}. The shaded regions are the phases chosen for
input into the more detailed atmospheric models. In this paper we only
consider models from the third shaded region (between $\sim$7000 and
8000 days), which we label the {\rm fx} series of models.

The grey radiative transfer in the outer layers of these models is
based on Rosseland opacities of \citet{Rogers92} and
\citet{Ferguson05}. In the outer atmosphere, the simple ad-hoc
modification of the radiative diffusion equation at small optical depths 
used by \cite{Fox82} is adopted. {\em This approximation
guarantees $T \propto r^{-0.5}$ in the outer layers.} However, this
approach is not accurate enough for spectrum computation, so we choose
to re-solve for the gas temperature, as described in the following
section. 

\subsection{The atmospheric models}

With the gas pressure fixed from the dynamical models, the gas and
dust temperatures in the photosphere are {\rm re-solved by 
using an opacity-sampling method in LTE}. This is in contrast to 
previous models of Mira atmospheres that used a small 72-wavelength
averaged-opacity mesh
\citep[e.g.][]{Hofmann98,Ireland04c,Ireland04d,Ireland06} or
51-wavelength mesh \citep[e.g.][]{Hofner03}. Our continuum and
Just-Overlapping-Line-Approximation (JOLA) VO opacities are the same as in
\citet{Hofmann98}. The continuum opacity sources (free-free and
bound-free where applicable) are H, H-, H$_2$, H$_2$-, He- and
Thompson scattering. The 
H$_2$O lines come from \citet{Partridge97} and those of TiO from 
\citet{Schwenke98}. Other diatomic molecular (CO, OH, CN, 
SiO, MgH) and neutral atomic lines (Na, Mg, Al, K, Ca, Ti, V, Cr, Mn, Fe,and
Ni) come from the
input to the ATLAS12 models \citep{Kurucz94}. The 
lack of metal bound-free opacities means that {\em our models are
inaccurate at wavelengths shortwards of 450\,nm}: wavelengths that are both
unimportant energetically in the atmospheric layers modelled by the
opacity-sampling code and strongly affected by numerous other
approximations such as LTE. 


A micro-turbulence of 2.8\,km/s is
assumed, in between that used in the M-giant models of \citet{Plez92}
and that derived by \citet{Hinkle79}. Line profiles for molecules are
taken to be  Gaussian, as pressure-broadening is negligible and the opacity is
dominated by a very large number of weak lines. For atomic lines, a
Voigt profile is assumed, with only radiative damping taken into
account with a nominal $\Gamma$ value of $10^8$, typical of
atomic transitions. 
The equation of state is based
on that used in \citet{Tsuji73}, with 35 atoms and their first 2
positive ionization 
states, 60 molecules, H-, H$_2$- and F-. The chemical equilibrium constants
for CO, MgH, SiH and TiO$_2$ have since been measured more
accurately and differ significantly from \citet{Tsuji73}, so for these
molecules we have adopted constants from 
\citet{Sharp90}. We assume solar abundances, and take the solar
abundances from \citet{Grevesse96}. 

The dust opacity and equation of state is taken from the
approximations of \citet{Ireland06}, with parameters designed to best
fit dust scattering opacity observations: the log of the number of nuclei per H
atom $\log(N_{\rm nuc}) = -12.8$ and the ratio of sticking
coefficients $\alpha_s=1$. This treatment of dust opacity is
temperature-dependent, due to the $\sim 1\mu$m dust opacity being a
strong function of the Fe-content of the dust, only approaching
standard interstellar medium dust opacities \cite[e.g.][]{Draine84} at
$T \la 1000K$.

The opacities are calculated for 4300 wavelengths running from 200\,nm to
50\,$\mu$m, with 1\,nm sampling between 200\,nm and 3\,$\mu$m and sampling
proportional to $\lambda^3$ for longer wavelengths. The effect of the velocity
stratification on the radiative transfer is not taken into account,
but is expected to have significant effects only in the vicinity of
shock fronts (see Section~\ref{sectQuality}).  

Instantaneous relaxation of shock
heating behind the moving shock front is assumed \citep[see
  discussion in][]{Bessell89}. This is equivalent to enforcing
$L=L_{\rm surf}$ everywhere, with $L_{\rm surf}$ the model surface
luminosity. Although the shock luminosity is important for modelling
emission lines, its luminosity is generally negligible in the
photospheric regions where the continuum is optically thin and
where lines
are formed. At phases immediately before maximum luminosity, as the
light curve of a Mira is rapidly rising, the shock luminosity can for
a small time reach half the model luminosity. The assumption $L=L_{\rm
  surf}$ is incorrect in this case, meaning that the shape of the
continuum during this rapid pre-maximum flux increase phase may not be a
good approximation. The effects of this approximation are examined
further in Section~\ref{sectOther}. 

The equation of radiative transfer is solved using the same code as
\citet{SchmidBurgk84}, with up to 80 discrete depth points. 
Since this code uses a spline interpolation, the pressure
discontinuity at the shock front is smoothed out over 0.05$R_p$ (deep 
layers) to 0.1$R_p$ (high layers). The model outer-boundary is fixed
at 5\,$R_p$: a boundary where the dust formation is more complex than our
simple approximations allow, and where radiative acceleration on dust
can influence the dynamics significantly. Layers outside 5\,$R_p$ can
be considered the `wind' zone, and can be seen observationally only in
the mid-infrared due to dust opacity and in the strongest of molecular transitions.

Once the equation of radiative transfer is solved, the spectrum and
center-to-limb profile is computed over a finer wavelength grid,
typically with 17000 wavelengths. This finer grid includes wavelength
in the radio regime, where the wavelength-dependence of free-free
opacities means that the continuum photospheric radius is quite
different to that in the near-infrared \cite{Reid07}.


\section{Discussion of model quality}
\label{sectQuality}

\subsection{Verification of opacity-sampling code for a static model}

Previously published models for static extended M-giant atmospheres
from other groups \citep{Fluks94,Hauschildt99b} have demonstrated
relatively good agreement with observations. Therefore, as
a first step in verifying the performance of our code, we will
compare the model spectra for a relatively compact static model to
those from the {\tt PHOENIX} code. This code has also adopted LTE for
published models \citep{Hauschildt99b}, so if the input opacities to
their models and our models were the same, one would expect the model
outputs to be the same. As a comparison model, we choose a
5 solar-mass model with log($g$)=0 (cgs units), and $T_{\rm eff}=3200$\,K. This model
is one of the most extended configurations from \citet{Hauschildt99b},
which is not designed for very extended atmospheres, 
and is the most compact model where our numerical approximations still
solve the equation of radiation transport with reasonable accuracy.

\begin{figure}
 \includegraphics[width={1.0\columnwidth}]{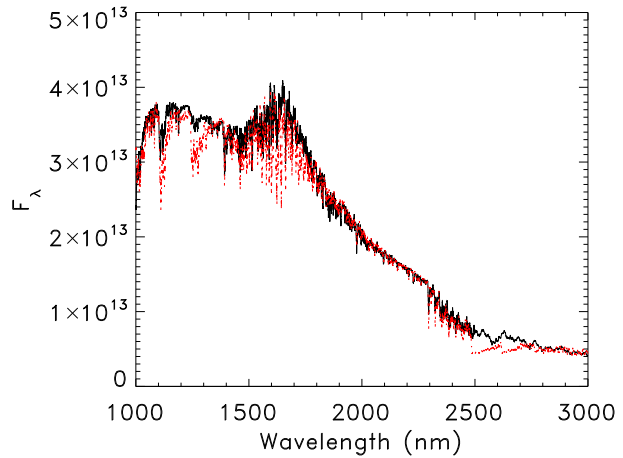}
 \caption{The spectrum between 1 and 3.0\,$\mu$m produced from the
 {\tt CODEX} atmosphere models at 3200\,K (solid
 lines), compared with that from the {\tt PHOENIX} atmosphere code
 (red dotted line). }
 \label{figPHOENIXCompIR}
\end{figure}

\begin{figure}
 \includegraphics[width={1.0\columnwidth}]{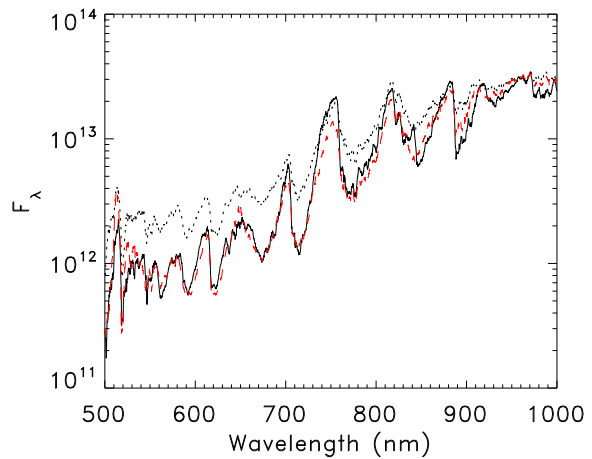}
 \caption{The spectrum between 0.5 and 1.0\,$\mu$m produced from the
 {\tt CODEX} atmosphere models at 3200\,K (solid
 lines), compared with that from the {\tt PHOENIX} atmosphere code
 (dotted line), and the {\tt CODEX} models using the same TiO line
 list as {\tt PHOENIX} (red dashed line) (see text).}
 \label{figPHOENIXCompVIS}
\end{figure}

Figure~\ref{figPHOENIXCompIR} shows the model spectra comparison between 1 and
3.0\,$\mu$m, and Figure~\ref{figPHOENIXCompVIS} shows the comparison
between 0.5 and 1\,$\mu$m. The most noticeable difference is the depth
of the TiO features: the dominant molecular bands short-wards of
1\,$\mu$m. Some of this difference comes from the use of the
\citet{Jorgensen94} line list by \citet{Hauschildt99b} (N.B. more
recent {\tt PHOENIX} models of M-dwarfs use the same 
TiO opacity as we do).
When the \citet{Jorgensen94} line list is used with the {\tt CODEX}
models \citep[adopting the oscillator strengths from][]{Allard00}, the
difference at around 750\,nm becomes smaller and the details of the
band shapes agree more, as seen in
Figure~\ref{figPHOENIXCompVIS}. The discrepancy in the TiO feature at
1.25\,$\mu$m as shown in Figure~\ref{figPHOENIXCompIR} is also reduced
using this line list
(not shown). However, the overall flux level of models
the at shorter wavelengths, particularly between 500 and 700\,nm remains discrepant. 
Both reducing the effective temperature of
our models by $\sim$100\,K and using the \citet{Jorgensen94} line list
produces a very small difference between the {\tt CODEX} and {\tt
  PHOENIX} models (not shown).

\begin{figure}
 \includegraphics[width={1.0\columnwidth}]{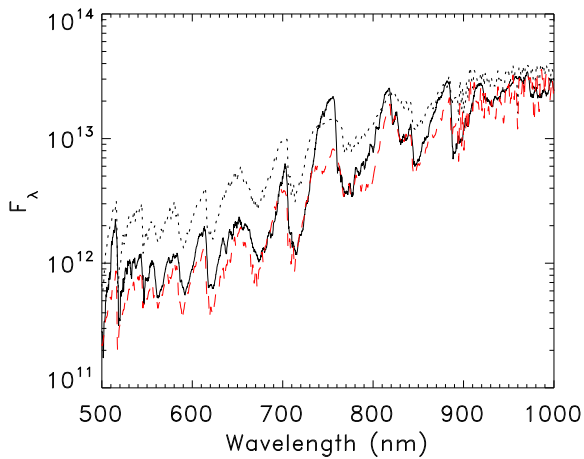}
 \caption{The spectrum between 0.5 and 1.0\,$\mu$m produced from the
 {\tt CODEX} atmosphere models at 3200\,K (solid
 lines), compared with an M6 giant spectrum (upper dotted line) and M7
 giant spectrum (lower red dashed line) from \citet{Fluks94}.}
 \label{figFluksComp}
\end{figure}

We also compare the 3200\,K static model to observations in
Figure~\ref{figFluksComp}, where an M6 and an M7 spectrum is compared
to the models.
We note that the difficulties in our models fitting the gap between
the strong TiO bands at 750\,nm is in common with the current version of the 
{\tt PHOENIX} models \citep[e.g. see][]{Lancon07}, due to an
inadequacy in the line lists.
Until this line list is updated, spectral indices based
on TiO absorption will not be able to be used to convert the model
spectra to spectral types. 

\begin{figure}
 \includegraphics[width={1.0\columnwidth}]{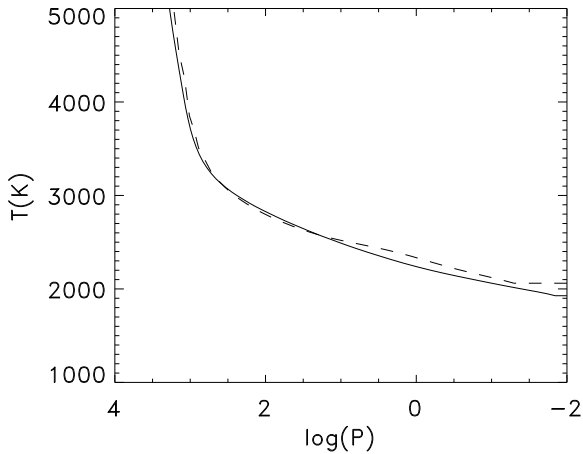}
 \caption{The temperature profile of the {\tt CODEX} models for a
 3200\,K model (solid line) plotted with the the {\tt PHOENIX} models (dashed
 line) for the same model parameters.}
 \label{figPHOENIXCompTemp}
\end{figure}

At $\sim2.5\mu$m, the {\tt PHOENIX} models have a
feature that is not in our models: presumably this is due to the
difference in water line lists: again the line list used by us is that
adopted in more modern {\tt PHOENIX} dwarf models and not the
published {\tt PHOENIX} giant models. There is also a
difference in the temperature structure 
of the models, as seen in Figure~\ref{figPHOENIXCompTemp}. We have
independently checked the temperature structure of our models using a
(computationally slow) Interactive Data Language ({\tt IDL}) radiative transfer
code that uses the a modified Uns\"old-Lucy temperature correction for
spherical atmospheres \cite[e.g.][]{Hauschildt99b} and trapezoidal-rule
integration, finding differences of order 10\,K only, so we therefore expect the
difference in temperature structure to be due to further opacity differences
rather than computational errors.

Note that previous versions of our code, such as that used in
\citet{Ireland04c,Ireland04d} could not reproduce the spectra of 
these static models nearly so well, due to the inaccuracies of the
Just Overlapping Line Approximation at these low to moderate TiO
optical depths \citep[e.g.][]{Brett90,Tej03}. 

\subsection{Critical assumptions examined at 3 model phases}
\label{sect3Mod}

There are a large number of computational checks one can make on the
quality of our atmospheric models before comparing them directly to
observations. The checks that we consider most relevant will be presented here,
relating directly to model physical assumptions and approximations: 1)
Is a grey temperature-profile adequate for describing the
atmospheric dynamics?; 2) Does the velocity stratification
significantly influence the solution to 
the radiative transfer equation?; 3) How much does the LTE
approximation affect the spectrum and temperature profile?; 
and 4) How does neglecting the shock luminosity by insisting
``$L=L_{\rm surf}$'' everywhere 
affect the model properties? We will attempt to
answer these questions by examining three model phases in detail. The
3 models have phases of 0.20 0.59 and 0.80 relative to (estimated)
maximum bolometric luminosity, which is assigned phase 0.0.

\begin{figure*}
 \includegraphics[width={1.0\textwidth}]{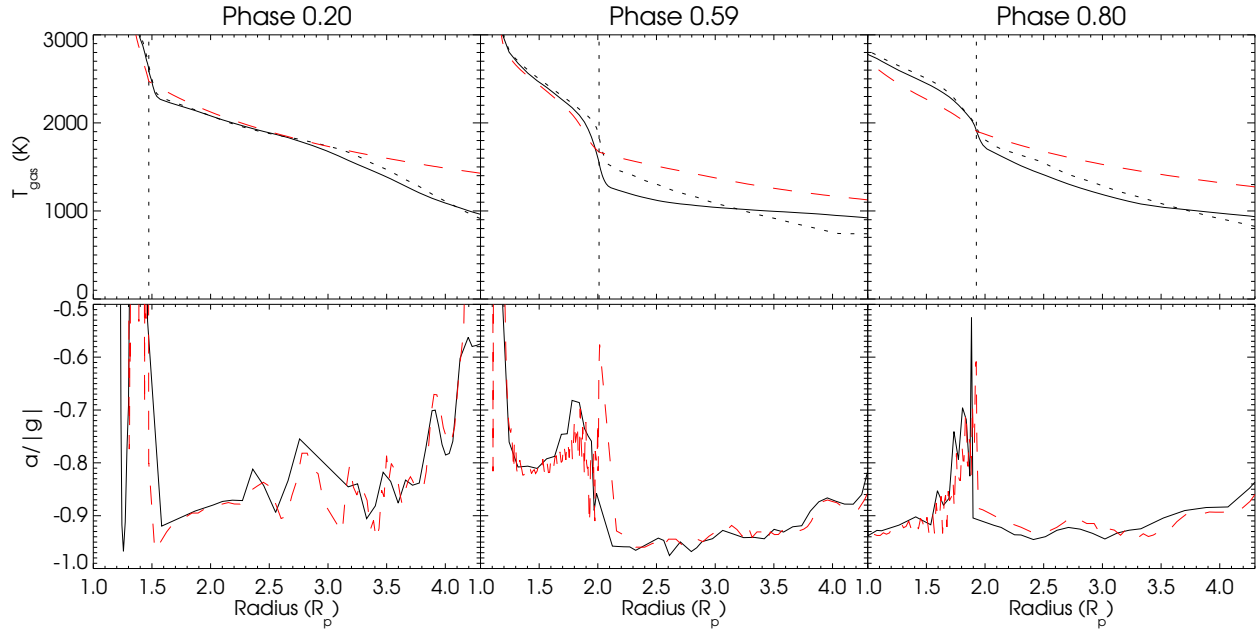}
 \caption{A comparison between the grey dynamical model temperature
 profile and total acceleration as a fraction of gravity
 (red dashed-lines) with that from the non-grey opacity-sampling code
 (solid lines), for the {\tt fx} series models at phases
 0.20 (left), 0.59 (center) and 0.80 (right) from the same cycle. The
 `spikes' in the acceleration  
 are near the location of shock fronts and are an artifact of
 smoothing the shock front in the dynamical grey and non-grey
 models. The dominant shock fronts are marked with vertical
 dotted lines, and the dotted lines in the top panel are temperatures
 calculated with the 72 mesh wavelength grid as in \citet{Ireland04c}.} 
 \label{figGreyComp}
\end{figure*}

Using a grey model for the non-linear dynamical models is only valid
to the extent that the temperature profile is consistent with that
from a more accurate non-grey model. Furthermore, the grey model uses
approximations for the spherically-extended atmosphere that also
affect the way that radiative acceleration is
calculated. Figure~\ref{figGreyComp} shows the grey and non-grey
temperature profiles, and the grey and non-grey accelerations as a
function of radius. The total acceleration includes a gravitational term, a
gas pressure term and a radiative acceleration term:

\begin{equation}
 a = g + \frac{1}{\rho} \frac{dP_{\rm gas}}{dr} + \frac{\pi}{c}
  \int_0^{\infty} k_{\nu} F_{\nu} d\nu.
\end{equation}

Here $k_\nu$ is the extinction coefficient at frequency $\nu$,
$F_{\nu}$ is the monochromatic flux per unit frequency and other terms
have standard meanings.

This figure demonstrates first that the grey approximation only
affects temperatures in 
the outer layers at the 10-20\% level (excluding shock fronts). A 10\%
increase in the grey model temperature would 
cause a 10\% increase in the pressure gradient for the same mass
stratification. This in turn changes the acceleration (both continuous
acceleration and the $\delta$-function acceleration at a shock-front)
at the 10\% level. During a pulsation cycle, the bulk of the
non-gravitational acceleration of a given layer occurs close to the
continuum-forming photosphere, where the diffusion approximation as
used in the grey model is most accurate. The gas in the outer layers
follows near-ballistic trajectories. Therefore, we
estimate that geometric pulsation amplitudes for layers between 1 and
4\,$R_p$ are accurate to $\la$10\%.


As a dynamic atmosphere has a velocity stratification $v(r)$, the
radiation field seen from a test point at a spectral-line wavelength may
strongly depend, as a consequence of Doppler shifting, on the depth of the 
layer of origin and on the direction of the line-of-sight. 
There are two classes of $v(r)$ effects: a geometric $v(r)$ projection
effect that is often moderate except for the very most extended
atmospheres; and more significant effects at and near a shock front 
where $v(r)$ is discontinuous at and has a strong gradient near the position of
the shock. Sample wavelengths that, e.g., are located at line frequencies 
within a forest of blanketing lines below the shock appear in the 
continuum above the shock, and vice versa. This may lead to
substantial errors in computed equilibrium temperature
when an opacity sampling technique is used for treating radiative transport 
that is dominated by line blanketing. Solutions for the case of a velocity 
stratification $v(r)$ have been developed for the opacity distribution
technique, though for only rather simple cases of $v(r)$ 
\citep[e.g.][]{Baschek01,Wehrse03,Castor04}, but no solutions are
known for the opacity sampling method (which necessarily has large
gaps in-between individual wavelengths that are sampled).

\begin{figure}
 \includegraphics[width={1.0\columnwidth}]{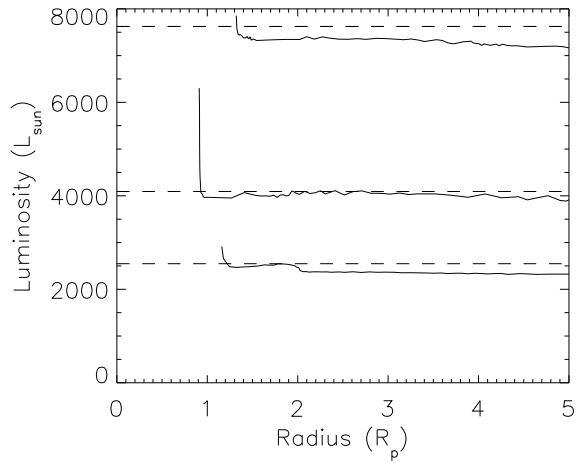}
 \caption{A comparison between the luminosity as a function of radius
 between the opacity-sampling code (straight dashed lines) and full
 dynamical modelling taking into account all wavelengths and the
 velocity stratification (solid lines). The models, from top to
 bottom, are at phases 0.20 (7623\,L$_\odot$),  0.80
 (4092\,L$_\odot$), and 0.59 (2548\,L$_\odot$).} 
 \label{figDynamicComp}
\end{figure}

\begin{figure*}
 \includegraphics[width={1.0\textwidth}]{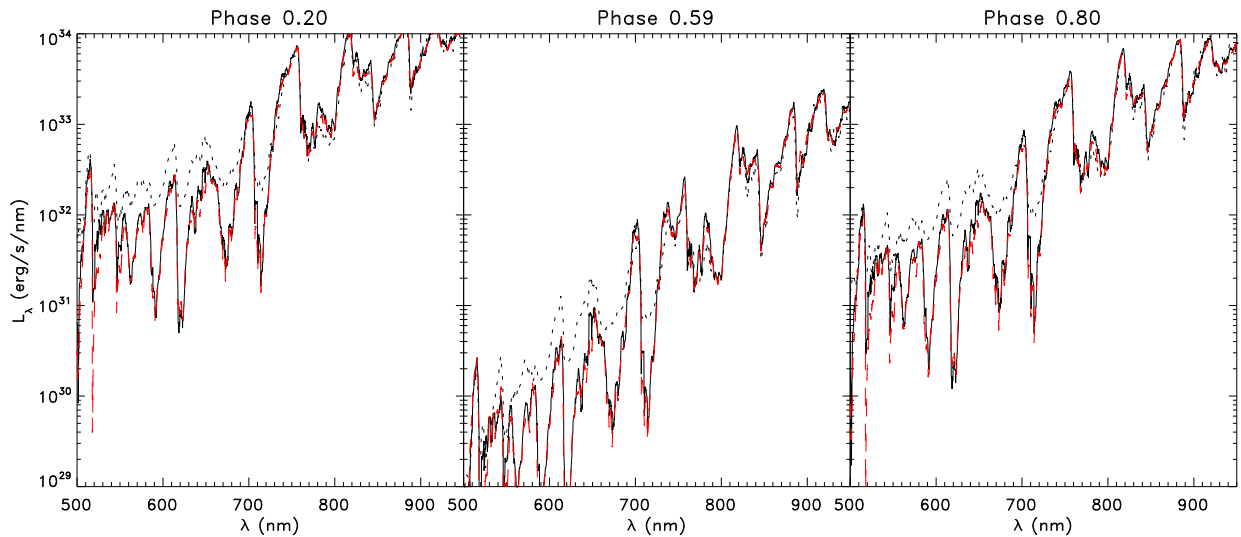}
 \caption{A comparison between the opacity-sampling code spectrum (solid lines)
 between 500 and 950\,nm and that resulting from a fluorescence
 scattering non-LTE treatment of the opacity (dotted lines). The
 dynamical atmosphere calculation is also shown (red dashed lines),
 demonstrating that this effect is very small compared with non-LTE
 effects.} 
 \label{figNLTETio}
\end{figure*}

We have, however, numerically examined the equation of radiative transfer 
in the dynamical model atmosphere without attempting a full
solution. This was accomplished by calculating the luminosity
at each layer of a model 
stratification on a grid of $\sim 10^6$ wavelengths, self-consistently
taking into account the velocity of each layer, by red-shifting or
blue-shifting the line profiles according to the difference in
projected velocities. This code was far too slow to attempt a solution
to the equation of radiative transfer, but could examine the departure
from the $L=L_{\rm surf}$ criterion. Figure~\ref{figDynamicComp} shows
a maximum 2\,\% deviation from the $L=L_{\rm surf}$ criterion,
demonstrating that the effects of the dynamical stratification are a
relatively minor effect in calculating the temperature structure. The
effects of the dynamical structure on the overall spectral shape was
also examined and also found to be minor, as shown in
Figure~\ref{figNLTETio}.

Deviations from LTE tend to increase with decreasing
densities and pressures. With typically very low pressures ($< 1$
dyn\,cm$^{-2}$) in the atmospheric regions of Mira variable stars
where spectral features are formed, these stars are a long way from
LTE. This is because collisions can not thermalise the level
populations of atoms and molecules. 
Non-LTE effects are most noticeable for electronic transitions of atoms and
TiO, where typical Einstein~A coefficients of $\sim10^7$s$^{-1}$ greatly
exceed collisional rates of order $10^2$\,s$^{-1}$. Note that this is
quite different to the CO molecule, where Einstein~A coefficients for
vibrational transitions are of
order $10^1$\,s$^{-1}$ \citep{Chandra96}. 

These non-LTE effects for
TiO are much more important in Miras than in static M giants, despite
static M giants also having Einstein~A coefficients much higher than
collisional rates. This is because static M giant atmospheres are not
nearly as extended as Miras, so the thin outer layers are much warmer for
the same continuum-forming temperature. In turn, this means that the
band depths calculated in LTE are much deeper for Miras, increasing
the potential for non-LTE effects to be important in modelling.

A complete band-model non-LTE treatment of TiO opacity would require a
complete new radiative transfer code (both codes used in this paper
tabulate opacities as a function of LTE temperature and
pressure). However, we have attempted to treat the TiO absorption with
a partial band-model non-LTE treatment in order to examine some
effects of the LTE approximation. 

We make an assumption that the electronic singlet and triplet ground
states for TiO are populated in thermal equilibrium (i.e. in both
rotational and vibrational equilibrium), and that transitions to and
from excited electronic states occur via scattering processes. Both
resonance and fluorescence scattering processes are included. This
assumption is reasonable because radiative transitions between vibrational
levels are dominated by direct ro-vibrational transitions for
radiation field temperatures of $\sim$1500\,K (applicable to the
region where the strong TiO features form). These ro-vibrational
transitions see a thermalised radiation field due to strong
mid-infrared transitions of other molecules (notably
H$_2$O). Radiative transitions between electronic excited states
(multi-photon absorption) is not as important here as it is for many
atomic states because of the relatively large gap between the ground
and first electronic excited state (i.e. in layers where the strong
features are 
formed, the molecule spends the vast majority of its time in the
electronic ground state). Collisions with other atoms and molecules are also
negligible in the regions where the strongest features are formed. 

The scattering process is added to the equation for the source
function as follows:

\begin{equation}
 S(\lambda) = \frac{\kappa(\lambda) B(\lambda)+ \int \sigma(\lambda,\lambda') J(\lambda') d\lambda'}
 {\kappa(\lambda) + \int \sigma(\lambda',\lambda)d\lambda'}.
\end{equation}

$S$, $B$, $J$ and $\kappa$ have their usual meanings of source
function, Planck function, mean intensity and
absorption coefficient (cross-section per unit mass). 
The isotropic scattering matrix $\sigma(\lambda,\lambda')$ describes
scattering from wavelength $\lambda'$ to wavelength $\lambda$, in
units of cm$^2$/g/nm. For resonant (i.e. non-fluorescent) scattering,
$\sigma(\lambda',\lambda) = \delta(\lambda-\lambda')\sigma(\lambda)$,
with $\delta$ the Dirac $\delta$-function.

The results of this calculation is shown in
Figure~\ref{figNLTETio}. It is clear that fluorescence scattering in
general dominates the radiative processes in the $V$-band, and that
interpretations such as the simplified LTE interpretation of
\citet{Reid97} offer a gross over-simplification of the physical
processes that create the visible light-curves of Mira variables. This
statement is true both in the strong absorption bands and in the
regions of weaker absorption in-between the bands. Note that a similar
fluorescence approach would have to be applied to VO in order to model
the 0.75 to 1.2 micron region more accurately.

The models still do not well-represent the deep TiO 
features, as will be discussed in Section~\ref{sectObs}. A possible
explanation for this discrepancy is inaccurate 
data for TiO$_2$, that could enable TiO to be removed from the gas at
higher temperatures \citep[e.g.][]{Ireland05,Ireland06}, or
non-equilibrium processes in the chemical reactions of TiO. An example
of such a non-equilibrium process is the lowering of the vibrational
temperature of TiO as the mid-infrared regions of the spectrum become
optically-thin and the justification for the fluorescence scattering
approximation above no longer work. This would drive the reaction
TiO+H$_2$O$\rightleftharpoons$TiO$_2$+H$_2$ to the right, removing TiO from the
gas. This kind of complexity is beyond the capabilities of the code
developed for this paper.

\subsection{Other model quality issues}
\label{sectOther}


The opacity-sampling method for radiative transfer 
calculations is only valid if a large enough number of wavelengths are
used.
For several representative test models with different stellar parameters (in 
particular different effective temperature), we have solved the radiative 
transfer equation based on the here adopted 4300-wavelength mesh, as well as 
meshes with 2152 and 8606 wavelength samples. Differences of temperatures never
exceeded a few tens of degrees in high layers for the 4300 vs. 2152 case, and 
were negligible for the 4300 vs. 8606 case. Typically, including dust
absorption that makes the atmospheric high-layer opacity "greyer",
diminishes 4300 vs. 2152 differences.  
We also increased strongly heavy-element abundances, i.e. the effects of line 
blanketing upon the temperature stratification, and found increasing and 
significant 4300 vs. 2152 differences whereas no 4300 vs. 8606 differences 
showed up. Thus, a 4300-wavelength mesh is well suited to cover even quite 
unfavorable cases of M-type Mira atmospheres. Temperature errors of 
the order of a few tens of degrees are smaller than other errors such as those 
caused by the smoothing of shock fronts, the dynamic stratification, or non-LTE
effects. We note that the opacity-sampling models of \citet{Hofner07}
with only 64 wavelength points are not expected to have sufficient
wavelength sampling to accurately reproduce the
temperature-profile. However, those models also used a 64-wavelength 
opacity-sampling method for the dynamic atmosphere computation, which may be
preferable to our grey dynamical models.

As seen in Figure~\ref{figGreyComp}, the temperature profiles of the
{\tt CODEX} models differ noticeably from the profiles computed from
the models of \cite{Ireland04c}, where the opacities were mainly
approximated by mean opacities over 72 wavelength bins and dust
formation was not included. We see the
current models as a major improvement over those models. This older
style of opacity computation was intermediate between the current
opacity sampling method and a grey atmosphere, as can be seen from the
temperature profile over the range 1000 to 2000\,K. At the lowest
temperatures ($\la1000$\,K), the {\tt CODEX} models are warmed more
than these older models because of the presence of dust.

In addition to limitations evident from the tests above, the
{\tt CODEX} models have clear limits where 
unknown details of complex heterogeneous dust formation become
important, and could not be used to model e.g. OH/IR 
stars. These approximations are discussed in detail in  
\citet{Ireland06}. Importantly, dust is assumed to form in
equilibrium, which is only applicable to the hottest ($\ga$1000\,K)
dust.

Finally, we have performed test calculations where the 
$L=L_{\rm surf}$ approximation is removed, and the shock luminosity
is added to the models over a smoothed region of size $\Delta r/r \sim
5$\%. The only test model with a noticeable (i.e. above numerical errors)
change in model structure was the phase 0.80 model, which had a
540\,$L_\odot$ shock within the continuum-forming photosphere 
and a total luminosity of 4090\,$L_\odot$. The model that included the
shock luminosity had 5\% less H-band flux and 10\% more V-band
flux. As this phase is at the time where the model luminosity (and
light curve of observed Miras)
is rapidly increasing prior to maximum luminosity, these differences
are not very significant.

\section{Reliable Observational Predictions}
\label{sectObs}

Given that Section~\ref{sectQuality} demonstrated that there are
some approximations that are not physically realistic, we will examine
which model outputs are most reliable for comparison to
observations. Clearly, only those regions of the spectrum that are
well-approximated by LTE should give reasonable agreement to
observations. The phase, cycle and parameter dependence of predictions will
be examined in a forthcoming paper II.

\subsection{Spectra}

\begin{figure}
\includegraphics[width={1.0\columnwidth}]{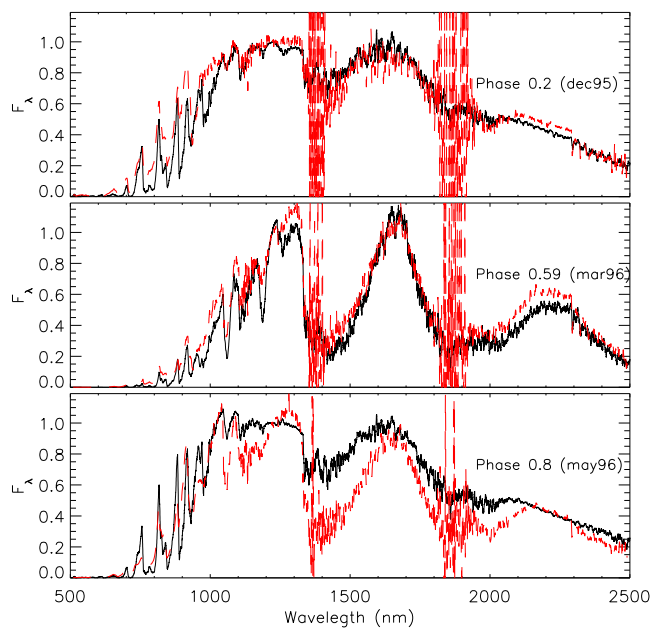}
\caption{Smoothed model spectra (black solid lines), with measured spectra of R~Cha from
  \citet{Lancon00} over-plotted (red dashed lines). The plots are labeled
  by their model phases. The estimated visual phases of the
  observations are 0.25, 0.5 and 0.76. } 
\label{figObsSpectra}
\end{figure}

Figure~\ref{figObsSpectra} shows some observed spectra compared with
models. Instead of spectra of $o$~Ceti, that were not readily available in
electronic form, we chose to compare the models with R~Cha, a Mira with
a 335 day period: almost exactly the same as $o$~Ceti.
For phases 0.2 and 0.59, an observed spectrum is available at
a close enough model phase to give reasonable spectral agreement
between the {\tt CODEX} models and observations. In particular, the
near-minimum spectrum is much 
better represented by the {\tt CODEX} model than by the previous generation of models
as examined in \citet{Tej03}. The infrared features of H$_2$O and CO
are reproduced reasonably, but the TiO features remain too deep in the
{\tt CODEX} models. Although some of this discrepancy is explained by
the effect of the LTE approximation, the test calculation results shown in 
Figure~\ref{figNLTETio} demonstrate that a fluorescence scattering
approximation does not noticeably affect band depths long-wards of
750\,nm. Therefore, we can not consider the band depths of strong TiO
features to be a reliable prediction, with or without the addition of
fluorescence scattering. The infrared features, particularly those
short-wards of 2500\,nm, are expected to be a reasonably good model
prediction. 

The visible-light spectrum and flux is too small to be seen on the
linear scale in Figure~\ref{figObsSpectra}. As visible light curves
are one of the richest data sets for Mira variables, theoretical
predictions for these curves as a function of physical parameters is a
major goal of our effort. As visible wavelengths are in the Wien part
of the Planck function at temperatures relevant to Miras, fluxes and
spectra are very sensitive to changes in the atmospheric stratification.
The effects of the LTE approximation as
discussed in Section~\ref{sect3Mod} cause visible fluxes to be
underestimated by a factor of $\sim$2. The difficulty in accurately
treating dust can give another few tens of \% error. So we expect the
model-predicted visible light curve to have uncertainties of
approximately 1 magnitude. 

\subsection{Diameters}

\begin{figure}
\includegraphics[width={1.0\columnwidth}]{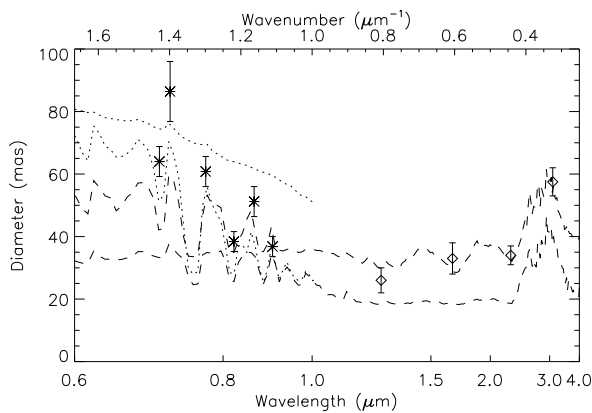}
\caption{Model diameters between 0.6 and 4\,$\mu$m for the {\tt CODEX}
  models (dashed lines) and the models with the fluorescence
  scattering approximation for TiO (dotted line). The
  range shown corresponds to the maximum and minimum diameters of the
  three models using two fitting techniques (see text). The horizontal
  axis is linear in wave-number so that both the full range and the
  TiO features can be shown. Points with error bars
  show the range of measured diameters from \citet{Ireland04a} and
  \citet{Woodruff08}.} 
\label{figMeasDiam}
\end{figure}

In Figure~\ref{figMeasDiam}, we show the range of model-predicted
diameters for $o$~Ceti for the three phases examined. In order to make
this plot, we calculated visibility curves for filters with 1\%
fractional bandwidth and calculated the best fit uniform-disk
visibility curves at $V=0.6$ and $V=0.2$. The fitted angular diameter
can be a strong function of where this fit is made
\cite[e.g][]{Ireland04c}. We assume a distance of 100\,pc
after \citet{vanLeeuwen07} for this 
plot. We also show measured diameters from \citet{Ireland04a}
($<1$\,$\mu$m) and \citet{Woodruff08} (1-4\,$\mu$m). We do not show
mid-infrared diameters from  \citet{Weiner03}, because of the
complexity of accounting for the over-resolved dust emission from the
wind. The error bars represent both observational error and
the differences amongst observations at different phases. The J-band
diameter range of the models match the observations very well, showing
that the continuum diameters are in good agreement: certainly within
10\% in diameter, corresponding to 5\% in effective temperature. However, the
measured diameters at H, K and L bands are always close to the largest
model diameters. We suspect this not to be an error in 
modelling, but instead a physical effect demonstrating that $o$~Ceti is
usually surrounded by more extended molecular layers than the present
models produce. Changing
e.g. the assumed model mass may help resolve this small discrepancy.

In \citet{Ireland06}, we showed that the diameters of Mira variables
at wavelengths shorter than 1\,$\mu$m were in general well-described
by the model of dust formation that we use here, consistent with
Figure~\ref{figMeasDiam}. However, this
statement was not true for the strong TiO absorption bands, in
particular the band at 712\,nm. The inclusion of the fluorescence
scattering approximation as a test calculation in this paper greatly
increases the model diameters in the TiO absorption bands (as seen by
the dotted line), meaning that, where necessary, more accurate
predictions for model diameters short-wards of 1\,$\mu$m can now be
produced. 

\section{Conclusions and Future Work}

We have presented the Cool Opacity-sampling Dynamic EXtended 
({\tt CODEX}) modelling method: an improved scheme for constructing stellar 
  atmosphere models that is tailored to Mira variable star
  atmospheres. This scheme includes self-excited
pulsation 
and a 4300-wavelength grid opacity-sampling code to solve for the
equation of radiative transfer. 

The models stand up to a variety of numerical tests, including
adequate treatment of shock fronts and dynamical effects, and
sensitivity at less than the 10\% to the grey approximation used 
in the dynamical models. The models notably have a $\sim$100\,K
maximum difference in temperature profile when benchmarked against
static {\tt PHOENIX} models, which give the {\tt CODEX} models deeper
absorption shortwards of 950\,nm. Test calculations using a
fluorescence scattering approximation for non-LTE TiO effects
demonstrated a modest difference in overall energetics, but a factor
of $\sim$2 increase in V-band flux and much larger diameters in TiO
absorption bands. 

Model predictions from the series presented here, as well as future
work, will be be made available
online\footnote{http://www.physics.usyd.edu.au/$\sim$mireland/codex/}. 
Work in progress includes the detailed analysis of the phase and cycle
dependence of model properties, models with different fundamental parameters 
including stars with longer periods and modified element abundances 
(metallicity, S-type C/O ratio, modified N abundance), as well as 
further comparison of typical model predictions with observed features.

\section*{Acknowledgments}
M.I. would like to acknowledge Michelson Fellowship support from the
Michelson Science Center and the NASA Navigator Program. M.S. would
like to acknowledge support from the Deutsche Forschungsgemeinschaft
with the grant "Time Dependence of Mira Atmospheres". 


\label{lastpage}

\end{document}